\documentclass{aa}
\usepackage{graphicx}
\usepackage{bm}
\usepackage{times}
\usepackage{amsmath}
\usepackage{amssymb}
\usepackage{natbib}
\usepackage{color}
\usepackage{mathrsfs}
\usepackage[colorlinks,allcolors=blue]{hyperref}
\usepackage{url}
\usepackage{multirow}
\bibpunct{(}{)}{;}{a}{}{,}
\graphicspath{{./fig/}{./png/}}
\usepackage{xcolor}

\newcommand{\uuu}{{\bm u}}

\newcommand{\xxx}{{\bm x}}

\newcommand{\FFF}{{\bm F}}

\newcommand{\gggg}{{\bm g}}
\newcommand{\SSt}{\bm{\mathsf{S}}}
\newcommand{\SStij}{\mathsf{S}_{ij}}

\newcommand{\Eqs}[2]{Equations~(\ref{#1}) and (\ref{#2})} 
\newcommand{\Equs}[2]{Equations~(\ref{#1}) to (\ref{#2})}
\newcommand{\Equsa}[2]{Eqs.~(\ref{#1}) to (\ref{#2})}
\newcommand{\EQ}{\begin{equation}}
\newcommand{\EN}{\end{equation}}
\newcommand{\EQA}{\begin{eqnarray}}
\newcommand{\ENA}{\end{eqnarray}}
\newcommand{\brac}[1]{\langle #1 \rangle}
\newcommand{\pd}{\partial}

\newcommand{\mean}[1]{\overline{#1}}

\newcommand{\cP}{c_{\rm P}}
\newcommand{\cV}{c_{\rm V}}
\newcommand{\cs}{c_{\rm s}}

\newcommand{\nutz}{\nu_{\rm t0}}

\newcommand{\urms}{u_{\rm rms}}

\newcommand{\ustar}{u_\star}

\newcommand{\AAV}{A_{\rm V}}
\newcommand{\AAH}{A_{\rm H}}

\newcommand{\lcorr}{\ell_{\rm corr}}

\newcommand{\chiSGS}{\chi_{\rm SGS}}

\newcommand{\Co}{{\rm Co}}

\newcommand{\CoF}{{\rm Co}_{\rm F}}

\newcommand{\Hp}{H_{\rm p}}

\newcommand{\Pe}{{\rm Pe}}

\newcommand{\PeSGS}{{\rm Pe}_{\rm SGS}}
\newcommand{\Pra}{{\rm Pr}}

\newcommand{\PraSGS}{{\rm Pr}_{\rm SGS}}

\newcommand{\Rat}{{\rm Ra}_{\rm t}}
\newcommand{\RaF}{{\rm Ra}_{\rm F}}
\newcommand{\RaFS}{{\rm Ra}_{\rm F}^\star}
\newcommand{\RaFc}{{\rm Ra}_{\rm F}^{\rm c}}

\newcommand{\Rey}{{\rm Re}}

\newcommand{\Recs}{{\rm Re}_{\rm c_s}}

\newcommand{\Ta}{{\rm Ta}}

\newcommand{\taucool}{\tau_{\rm cool}}

\newcommand{\qij}{Q_{ij}}
\newcommand{\qxx}{Q_{xx}}
\newcommand{\qyy}{Q_{yy}}
\newcommand{\qzz}{Q_{zz}}
\newcommand{\qxy}{Q_{xy}}
\newcommand{\qxz}{Q_{xz}}
\newcommand{\qyz}{Q_{yz}}
\newcommand{\tQij}[1]{\widetilde{Q}_{#1}}

\newcommand{\qrt}{Q_{r\theta}}
\newcommand{\qrp}{Q_{r\phi}}
\newcommand{\qtp}{Q_{\theta\phi}}

\newcommand{\LamV}{\Lambda_{\rm V}}

{}

\newcommand{\calC}{{\cal C}}

\newcommand{\calR}{{\cal R}}

\newcommand{\nmax}{n_{\rm max}}
\newcommand{\Fbot}{F_{\rm bot}}

\newcommand{\Fn}{\mathscr{F}_{\rm n}}

\newcommand{\tbot}{{\rm bot}}

\newcommand{\zbot}{z_{\rm bot}}

\def\onethird{{\textstyle{1\over3}}}

\def\onehalf{{\textstyle{1\over2}}}
\def\onefifth{{\textstyle{1\over5}}}

\def\twofifteenths{{\textstyle{2\over15}}}
\newcommand{\Figa}[1]{Fig.~\ref{#1}}
\newcommand{\Figas}[2]{Figs.~\ref{#1} and \ref{#2}}

\newcommand{\Fig}[1]{Figure~\ref{#1}} 
\newcommand{\Figu}[1]{Figure~\ref{#1}}

\newcommand{\Sec}[1]{Section~\ref{#1}} 

\newcommand{\Table}[1]{Table~\ref{#1}}

\definecolor{ForestGreen}{RGB}{34,139,34}
\definecolor{AGray}{rgb}{.4,.4,.4}
\definecolor{LightYellow}{rgb}{1.,1.,.8}
\definecolor{LightCyan}{rgb}{0.88,1,1}
\begin{document}

\authorrunning{K\"apyl\"a}
\titlerunning{$\Lambda$ effect in convection}

   \title{$\Lambda$ effect in rotating hydrodynamic convection}

   \author{P. J. K\"apyl\"a
          \inst{1}
          }

   \institute{Institute for Solar Physics (KIS), Georges-K\"ohler-Alle
     401a, 79110 Freiburg im Breisgau, Germany, email:
     \href{mailto:pkapyla@leibniz-kis.de}{pkapyla@leibniz-kis.de}}

\date{\today}

\abstract{%
  Rotating anisotropic convection generates differential rotation in
  stellar convection zones.}%
{The main aim is to compute the non-diffusive contribution ($\Lambda$
  effect) to angular momentum transport, described by Reynolds stress,
  from rotating turbulent convection.}%
{Rotating hydrodynamic convection is simulated in Cartesian geometry
  at different latitudes and rotation rates. Large-scale flows are
  suppressed such that the Reynolds stress is due to non-diffusive
  effects.}%
{The radial angular momentum flux is downward (outward) for slow
  (fast) rotation. This is in contrast in prevailing theories in
  mean-field hydrodynamics where the radial transport is always
  downward. The outward transport at rapid rotation is due to thermal
  Rossby waves that manifest as elongated large-scale convection cells
  near the equator. The horizontal angular momentum flux is always
  equatorward, with increasing concentration toward the equator as in
  earlier Cartesian studies. The magnitudes of the $\Lambda$ effect
  coefficients are roughly an order of magnitude lower than in the
  case of anisotropically forced turbulence or in analytic theories.}%
{The current results highlight the tension between numerical
  simulations, widely used mean-field models, and solar observations.
  The mean-fields models have been remarkably successful in
  reproducing solar differential rotation but underlying assumptions
  regarding turbulence in these models seem to be at odds with 3D
  simulations. The current simulation results for the vertical
  (radial) angular momentum transport are in accordance with spherical
  shell simulations, where thermal Rossby waves are responsible for
  the generation of equatorial acceleration or solar-like differential
  rotation. Thermal Rossby waves are absent in the turbulence models
  of current mean-field theories and they have not been unambiguously
  detected in the Sun.}%


  \maketitle


\section{Introduction}

Observations of the Sun and other stars with convective envelopes show
that differential rotation is ubiquitous
\citep[e.g.][]{Rudiger_et_al_2013_Magnetic_Processes_in_Astrophysics,
  Reinhold_Gizon_2015_AA_583_65}. The main driver of differential
rotation has long thought to be non-vanishing Reynolds stress due to
the interaction of rotation and convective turbulence
\citep[e.g.][]{Rudiger_1989_Differential_Rotation_and_Stellar_Convection}
\begin{eqnarray}
\qij = \mean{u_i u_j},
\end{eqnarray}
where $u_i = U_i - \mean{U}_i$ is the fluctuating velocity and where
the overbar denotes a suitably defined average. More specifically, the
off-diagonal components of $\qij$ have non-diffusive contributions in
anisotropic turbulence that are proportional to the angular velocity
$\Omega$ which lead to generation of differential rotation. This is
more commonly known as the $\Lambda$ effect
\citep[e.g.][]{Rudiger_1980_GAFD_16_239}.

The $\Lambda$ effect has been studied with simulations of
anisotropically forced turbulence
\citep{Kapyla_Brandenburg_2008_AA_488_9,
  Kapyla_2019_AA_622_195,Kapyla_2019_AN_340_744,
  Barekat_et_al_2021_AA_655_79}. These results have confirmed the
existence of the $\Lambda$ effect and are in broad agreement with
mean-field hydrodynamics
\citep[][]{Rudiger_1989_Differential_Rotation_and_Stellar_Convection}. Convection
simulations show qualitatively similar results at slow rotation
\citep[e.g.][]{Pulkkinen_et_al_1993_AA_267_265,
  Rudiger_et_al_2005_AN_326_315, Rudiger_et_al_2019_AA_630_109,
  Hupfer_et_al_2005_AN_326_223, Rudiger_Kuker_2021_AA_649_173}, but
differences arise at rapid rotation where, for example, the radial
angular momentum flux changes sign
\citep{Chan_2001_ApJ_548_1102,Kapyla_et_al_2004_AA_433_793}. The
latter arises due to large-scale wave-like convective modes also known
as Busse columns, banana cells, or thermal Rossby waves that appear as
tilted columns in an equatorial latitude belt
\citep[e.g.][]{Busse_2002_PhFl_14_1301,
  Aurnou_et_al_2007_Icarus_190_110}; see \cite{Kapyla_2023_AA_669_98}
and \cite{Mori_Hotta_2023_MNRAS_519_3091} for recent analyses of the
associated angular momentum transport. Mean-field theories of stellar
angular momentum transport
\citep[e.g.][]{Kitchatinov_Rudiger_1995_AA_299_446,
  Kitchatinov_Rudiger_2005_AN_326_379,
  Kleeorin_Rogachevskii_2006_PRE_73_046303} do not include such
large-scale convective modes and therefore the accuracy of their
predictions in the case of convection is uncertain. The main goal of
the current study is to compute the $\Lambda$ effect from convection
using a comprehensive set of simulations at varying rotation rates for
the first time.

The paper is organized as follows: \Sec{sec:MFL} introduces the
$\Lambda$ effect in the framework of mean-field hydrodynamics,
\Sec{sec:model} describes the model. In \Sec{sec:results} the results
of the study are discussed and in \Sec{sec:conclusions} the
conclusions are presented.

\section{Mean-field theory and the $\Lambda$ effect}
\label{sec:MFL}

The Reynolds stress in rotating turbulence can
be written as
\begin{eqnarray}
\qij = \qij^{(0)} + \qij^{(\Omega)} + \qij^{(S)},
\end{eqnarray}
where $\qij^{(0)}$ is present in the absence of rotation or shear,
$\qij^{(\Omega)}$ is due to rotation, and $\qij^{(S)}$ is due to
shear. Assuming the large-scale fields vary slowly in space and time,
we write an ansatz
\begin{eqnarray}
\qij = \qij^{(0)} + \Lambda_{ijk} \Omega_k - \mathcal{N}_{ijkl}\frac{\pd \mean{U}_k}{\pd x_l},
\end{eqnarray}
where $\Lambda_{ijk}$ and $\mathcal{N}_{ijkl}$ are third and fourth
rank tensors, respectively, and $\bm\Omega$ is the mean angular
velocity.  The appearance of cross-correlations requires two preferred
directions
\citep[e.g.][]{Rudiger_1989_Differential_Rotation_and_Stellar_Convection}.
The gravity ($\gggg$) and angular velocity ($\bm\Omega$) vectors
fulfill this requirement in rotating convection. However, in such
cases large-scale flows are produced due to the $\Lambda$ effect, and
disambiguation of $\qij^{(\Omega)}$ and $\qij^{(S)}$ is impossible
using a single simulation. To circumvent this complication the mean
flows can be removed artificially
\citep[e.g.][]{Rudiger_et_al_2019_AA_630_109,Barekat_et_al_2021_AA_655_79}. Thus
$\qij^{(S)}$ vanishes, and the Reynolds stress reads
\begin{eqnarray}
\qij = \qij^{(0)} + \qij^{(\Omega)},
\end{eqnarray}
where the rotation-generated stress is
\begin{eqnarray}
\qij^{(\Omega)} = \qij- \qij^{(0)}.
\end{eqnarray}
The non-diffusive stress corresponding to the $\Lambda$ effect is then
\begin{eqnarray}
\qij^{(\Omega)} =  \Lambda_{ijk} \Omega_k,
\end{eqnarray}
which can be extracted from a single experiment because off-diagonal
components of $\qij^{(0)}$ vanish. The horizontal, meridional, and
vertical components of the $\Lambda$ effect are given by
\begin{eqnarray}
\qtp^{(\Omega)} &=& \nutz \Omega H \cos\theta, \label{equ:Qtp} \\
\qrt^{(\Omega)} &=& \nutz \Omega M \sin\theta\cos\theta, \label{equ:Qrt} \\
\qrp^{(\Omega)} &=& \nutz \Omega V \sin\theta, \label{equ:Qrp}
\end{eqnarray}
where $\nutz$ is a reference value of turbulent viscosity. The factor
$\nutz \Omega$ is used for normalization, whereas $V$, $H$, and $M$
are dimensionless position-dependent functions.

It is important to bear in mind that \Equs{equ:Qtp}{equ:Qrp} are
parametrizations that derive from symmetry properties but take no
stand on the physical origin of the stress. Furthermore,
\Equs{equ:Qtp}{equ:Qrp} are not limited to low Reynolds numbers which
is the case for results derived under the first-order smoothing
approximation (FOSA).

\section{The model} \label{sect:model}
\label{sec:model}

The model is the same as in several earlier studies \citep[e.g.][and
  references therein]{Kapyla_2024_AA_683_221}. The equations for
compressible hydrodynamics
\begin{eqnarray}
\frac{D \ln \rho}{D t} &=& -\bm\nabla \bm\cdot \uuu, \label{equ:dens}\\
\frac{D\uuu}{D t} &=& {\bm g} -\frac{1}{\rho}(\bm\nabla p - \bm\nabla \bm\cdot 2 \nu \rho \bm{\mathsf{S}}) - 2\bm\Omega\times\uuu,\label{equ:mom} \\
T \frac{D s}{D t} &=& -\frac{1}{\rho} \left[\bm\nabla \bm\cdot \left(\FFF_{\rm rad} + \FFF_{\rm SGS}\right)  +  \calC \right] + 2 \nu \bm{\mathsf{S}}^2,
\label{equ:ent}
\end{eqnarray}
were solved, where $D/Dt = \pd/\pd t + \uuu\bm\cdot\bm\nabla$ is the
advective derivative, $\rho$ is the density, $\uuu$ is the velocity,
$\bm{g}=-g\hat{\bm{e}}_z$ is the acceleration due to gravity with
$g>0$, $p$ is the pressure, $\nu$ is the constant kinematic viscosity,
$\SSt$ is the traceless rate-of-strain tensor with
$\SStij = \onehalf (u_{i,j} + u_{j,i}) - \onethird \delta_{ij}
\bm\nabla\cdot\uuu$ and
$\bm\Omega = \Omega_0(-\sin\theta,0,\cos\theta)^{\rm T}$ is the
rotation vector, where $\theta$ is the angle between $\bm\Omega$ and
$\gggg$. $T$ is the temperature, $s$ is the specific entropy,
$\FFF_{\rm rad}$ and $\FFF_{\rm SGS}$ are the radiative and turbulent
subgrid scale (SGS) fluxes, respectively, and $\calC$
describes surface cooling. The gas is assumed optically thick and
fully ionized such that radiation is modeled via diffusion
approximation. The ideal gas equation of state
$p= (\cP - \cV) \rho T =\calR \rho T$ applies, where $\calR$ is the
gas constant, and $c_{\rm P}$ and $c_{\rm V}$ are the specific heats
at constant pressure and volume, respectively. The radiative flux is
$\FFF_{\rm rad} = -K\bm\nabla T$, where $K$ is the radiative heat
conductivity given by $K = 16 \sigma_{\rm SB} T^3/(3 \kappa \rho)$,
where $\sigma_{\rm SB}$ is the Stefan-Boltzmann constant and $\kappa$
is the opacity. The Kramers opacity law
$\kappa =\kappa_0 (\rho/\rho_0)^a (T/T_0)^b$ with $a=1$ and $b=-7/2$
was used \citep{Weiss_et_al_2004_Cox_Giuli_Principles}; see also
\cite{Edwards_1990_MNRAS_242_224},
\cite{Brandenburg_et_al_2000_Astrophysical_Convection_and_Dynamos_85},
and \citep{Dorch_Nordlund_2001_AA_365_562} for early applications to
convection simulations.

Turbulent SGS diffusivity is applied for the entropy fluctuations with
$\FFF_{\rm SGS} = -\rho T \chiSGS \bm\nabla s'$, where
$s'(\xxx)=s(\xxx)-\mean{s}(z)$ and where the overbar indicates
horizontal averaging. $\chiSGS$ is constant throughout the domain and
$\mean{\FFF}_{\rm SGS} \approx 0$ because the SGS diffusivity operates
on the entropy fluctuations. The surface cooling is given by
$\calC = f(z) \cP \rho (T - T_{\rm cool})/\taucool$, where
$\taucool = 0.4 \sqrt{d/g}$ is a cooling timescale, $T=e/c_{\rm V}$ is
the temperature where $e$ is the internal energy, and where
$T_{\rm cool}=T_{\rm top}$ corresponds to the fixed value at the upper
boundary.

\subsection{Geometry, initial and boundary conditions}

The computational domain is a rectangular box where
$z_{\rm bot} \leq z \leq z_{\rm top}$ is the vertical coordinate, with
$z_{\rm bot}/d=-0.45$, $z_{\rm top}/d=1.05$, and where $d$ is the
depth of the initially isentropic layer (see below). The horizontal
coordinates $x$ and $y$ are given by $-2d \leq (x,y) \leq 2d$. The
initial stratification consists of three layers. The two lower layers
are polytropic with polytropic indices $n_1=3.25$
($z_{\rm bot}/d \leq z/d < 0$) and $n_2=1.5$ ($0 \leq z/d \leq
1$). The uppermost layer above $z/d=1$ is initially isothermal. The
initial velocity follows a Gaussian-noise distribution with amplitude
on the order of $10^{-4}\sqrt{dg}$.

The horizontal boundaries are periodic. The vertical boundaries are
impenetrable and stress-free for the flow such that
$u_{x,z}= u_{y,z}= u_z = 0.$ The temperature gradient at the bottom
boundary is given by $\pd_z T = -F_\tbot/K_\tbot,$ where $F_{\rm bot}$
is the fixed input flux and $K_\tbot(x,y,\zbot)$ is the heat
conductivity at $z=\zbot$. Constant temperature $T=T_{\rm top}$ is
assumed on the top boundary.

\subsection{Units, control parameters, and simulation strategy}

The units of length, time, density, and entropy are given by
$[x] = d$, $[t] = \sqrt{d/g}$, $[\rho] = \rho_0$, $[s] = \cP$, where
$\rho_0$ is the initial value of density at $z=z_{\rm top}$. The
profile $f(z)=1$ above $z/d=1$ and $f(z)=0$ below $z/d=1$, connecting
smoothly over a width of $0.025d$, and
$\xi_0=\Hp^{\rm top}/d = \mathcal{R}T_{\rm top}/gd$ sets the initial
pressure scale height at the surface. The Prandtl number based on the
radiative heat conductivity is $\Pr(\xxx,t) = \nu/\chi(\xxx,t)$, where
$\chi(\xxx,t)=K(\xxx,t)/\cP \rho(\xxx,t)$. SGS Prandtl number is given
by $\PraSGS = \nu/\chi$.

Rayleigh number based on the energy flux is given by
\begin{eqnarray}
\RaF = \frac{gH^4 \Fbot}{\cP \rho T \nu\chi^2}.
\end{eqnarray}
A diffusion-free Rayleigh number can be constructed using
$\Recs = \cs H/\nu$, via
\begin{eqnarray}
\RaFc = \frac{\RaF \Pra^2}{\Recs^3} = \frac{gH\Fbot}{\cP \rho T \cs^3}. \label{equ:RaFc}
\end{eqnarray}
Assuming $H= H_T \equiv \cP T /g$, $\RaFc$ reduces to the
dimensionless normalized flux
\citep[e.g.][]{Brandenburg_et_al_2005_AN_326_681}:
\begin{eqnarray}
\Fn = \Fbot/\rho \cs^3.\label{equ:Fn}
\end{eqnarray}
In rotating simulations, a flux-based diffusion-free modified Rayleigh
number is given by \citep[e.g.][]{Christensen_2002_JFM_470_115,
  Christensen_Aubert_2006_GeJI_166_97, Kapyla_2024_AA_683_221}
\begin{eqnarray}
\RaFS = \frac{\RaF}{\Pra^2 \Ta^{3/2}},\label{equ:RaFS}
\end{eqnarray}
where $\Ta= 4 \Omega_0^2 d^4 /\nu^2$ is the Taylor number. This can be
recast into a flux-based Coriolis number
\citep[][]{Kapyla_2024_AA_683_221, Bekki_2025_AA_703_262}
\begin{eqnarray}
\CoF = \frac{2\Omega_0 H}{\ustar} = 2\Omega_0 H \left(\frac{\rho}{\Fbot}\right)^{1/3} = (\RaFS)^{-1/3},
\end{eqnarray}
with $\ustar \equiv (\Fbot/\rho)^{1/3}$ and $H = H_T$.

In addition to explicit diffusion terms, the advective terms in
\Equs{equ:dens}{equ:ent} are written in terms of a fifth-order
upwinding derivative with a hyperdiffusive sixth-order correction with
a flow-dependent diffusion coefficient; see Appendix~B of
\cite{Dobler_et_al_2006_ApJ_638_336}. The {\sc Pencil
  Code}\footnote{\href{https://pencil-code.org/}{https://pencil-code.org/}}
was used to produce the simulations
\citep[][]{Pencil_Code_Collaboration_2021_JOSS_6_2807}.

\subsection{Diagnostics quantities}

The global Reynolds, SGS P\'eclet, and Coriolis numbers
\begin{eqnarray}
\Rey = \frac{\urms}{\nu k_1},\ \ \ \Pe_{\rm SGS} = \PraSGS\Rey,\ \ \ \Co = \frac{2\Omega_0}{\urms k_1},
\end{eqnarray}
describe the importance of viscosity, SGS diffusion and rotation
relative to advection, respectively. Here $\urms$ is the volume
averaged rms-velocity and $k_1=2\pi/d$ is an estimate of the largest
eddies in the system. Typically $\chi\ll\chiSGS$ in the convection
zone in the current simulations.

The overall anisotropy of the flow is characterized by the parameters
\begin{eqnarray}
\AAV = \frac{\qxx+\qyy-2\qzz}{Q}, \ \ \AAH = \frac{\qyy-\qxx}{Q},
\end{eqnarray}
where $Q = {\rm Tr} \qij$. The parameters $\AAV$ and $\AAH$ do not
bear information about the scale dependence of anisotropy. This is
revealed by spectral anisotropy parameters which are based on the power
spectra of velocity components \citep[e.g.][]{Kapyla_2019_AN_340_744}
\begin{eqnarray}
  \AAV(k) &=& \frac{E_x(k)+E_y(k)-2E_z(k)}{E_{\rm K}(k)}, \label{equ:AVk} \\
  \AAH(k) &=& \frac{E_y(k)-E_x(k)}{E_{\rm K}(k)}, \label{equ:AHk}
\end{eqnarray}
where $E_i \equiv u_i^2 = \int E_i(k) dk$, and
$E_{\rm K}(k) = \sum_i E_i(k)$. Diagnostics are time-averaged over the
statistically steady part of the simulations and horizontal or volume
averages are additionally applied. Error estimates are given as the
mean error of the mean where the number of turnover times is taken as
the number of realizations.

\begin{table}[t!]
\centering
\caption[]{Summary of the runs.}
  \label{tab:runs1}
       \vspace{-0.5cm}
      $$
          \begin{array}{p{0.05\linewidth}ccccccccccc}
          \hline
          \hline
          \noalign{\smallskip}
Run  & \RaFS  & \Ta [10^6]  & \theta & \CoF & \Co  & \Rey  & \Rat [10^6]  & \\
\hline
A1   &  2.5\cdot 10^{2}  & 0.01  &  0\degr &  0.28  &   0.07  &   38.9  &    4.1 \\
A2   &  2.5\cdot 10^{2}  & 0.01  & 15\degr &  0.28  &   0.07  &   38.9  &    4.1 \\
A3   &  2.5\cdot 10^{2}  & 0.01  & 30\degr &  0.28  &   0.07  &   38.9  &    4.2 \\
A4   &  2.5\cdot 10^{2}  & 0.01  & 45\degr &  0.28  &   0.07  &   39.0  &    4.1 \\
A5   &  2.5\cdot 10^{2}  & 0.01  & 60\degr &  0.28  &   0.06  &   39.0  &    4.2 \\
A6   &  2.5\cdot 10^{2}  & 0.01  & 75\degr &  0.28  &   0.07  &   38.9  &    4.2 \\
A7   &  2.5\cdot 10^{2}  & 0.01  & 90\degr &  0.28  &   0.07  &   38.9  &    4.0 \\
\hline
B1   &  3.1\cdot 10^{1}  & 0.04  &  0\degr &   0.55  &   0.13  &   39.4  &    4.3 \\
B1   &  3.1\cdot 10^{1}  & 0.04  & 15\degr &   0.55  &   0.13  &   39.4  &    4.3 \\
B3   &  3.1\cdot 10^{1}  & 0.04  & 30\degr &   0.56  &   0.13  &   39.5  &    4.3 \\
B4   &  3.1\cdot 10^{1}  & 0.04  & 45\degr &   0.56  &   0.13  &   39.7  &    4.3 \\
B5   &  3.1\cdot 10^{1}  & 0.04  & 60\degr &   0.56  &   0.13  &   39.8  &    4.2 \\
B6   &  3.1\cdot 10^{1}  & 0.04  & 75\degr &   0.57  &   0.13  &   39.7  &    4.2 \\
B7   &  3.1\cdot 10^{1}  & 0.04  & 90\degr &   0.57  &   0.13  &   39.8  &    4.2 \\
\hline
C1   &  3.9\cdot 10^{0}  & 0.16  &  0\degr &   1.04  &   0.26  &   39.6  &    4.5 \\
C2   &  3.9\cdot 10^{0}  & 0.16  & 15\degr &   1.05  &   0.25  &   39.8  &    4.5 \\
C3   &  3.9\cdot 10^{0}  & 0.16  & 30\degr &   1.06  &   0.25  &   40.0  &    4.4 \\
C4   &  3.9\cdot 10^{0}  & 0.16  & 45\degr &   1.06  &   0.25  &   40.3  &    4.5 \\
C5   &  3.9\cdot 10^{0}  & 0.16  & 60\degr &   1.07  &   0.25  &   40.4  &    4.4 \\
C6   &  3.9\cdot 10^{0}  & 0.16  & 75\degr &   1.10  &   0.25  &   40.5  &    4.4 \\
C7   &  3.9\cdot 10^{0}  & 0.16  & 90\degr &   1.11  &   0.25  &   40.6  &    4.4 \\
\hline
D1   &  2.5\cdot 10^{-1}  & 1.0  &  0\degr &   2.33  &   0.64  &   39.6  &    5.1 \\
D2   &  2.5\cdot 10^{-1}  & 1.0  & 15\degr &   2.35  &   0.63  &   39.9  &    5.1 \\
D3   &  2.5\cdot 10^{-1}  & 1.0  & 30\degr &   2.35  &   0.62  &   40.6  &    5.1 \\
D4   &  2.5\cdot 10^{-1}  & 1.0  & 45\degr &   2.35  &   0.62  &   41.0  &    5.2 \\
D5   &  2.5\cdot 10^{-1}  & 1.0  & 60\degr &   2.39  &   0.61  &   41.4  &    5.1 \\
D6   &  2.5\cdot 10^{-1}  & 1.0  & 75\degr &   2.49  &   0.60  &   42.0  &    5.0 \\
D7   &  2.5\cdot 10^{-1}  & 1.0  & 90\degr &   2.57  &   0.60  &   42.5  &    4.9 \\
\hline
E1   &  3.1\cdot 10^{-2}  & 4.0  &  0\degr &   4.56  &   1.27  &   39.8  &    6.0 \\
E2   &  3.1\cdot 10^{-2}  & 4.0  & 15\degr &   4.56  &   1.27  &   40.0  &    6.0 \\
E3   &  3.1\cdot 10^{-2}  & 4.0  & 30\degr &   4.53  &   1.25  &   40.5  &    6.1 \\
E4   &  3.1\cdot 10^{-2}  & 4.0  & 45\degr &   4.50  &   1.23  &   41.3  &    6.2 \\
E5   &  3.1\cdot 10^{-2}  & 4.0  & 60\degr &   4.55  &   1.19  &   42.5  &    6.3 \\
E6   &  3.1\cdot 10^{-2}  & 4.0  & 75\degr &   4.72  &   1.14  &   44.6  &    6.2 \\
E7   &  3.1\cdot 10^{-2}  & 4.0  & 90\degr &   4.89  &   1.11  &   45.8  &    6.1 \\
\hline
F1   &  3.9\cdot 10^{-3}  &  16  &  0\degr &   9.16  &   2.62  &   38.6  &    7.6 \\
F2   &  3.9\cdot 10^{-3}  &  16  & 15\degr &   9.16  &   2.61  &   38.9  &    7.6 \\
F3   &  3.9\cdot 10^{-3}  &  16  & 30\degr &   9.16  &   2.57  &   39.5  &    7.7 \\
F4   &  3.9\cdot 10^{-3}  &  16  & 45\degr &   9.22  &   2.51  &   40.4  &    7.8 \\
F5   &  3.9\cdot 10^{-3}  &  16  & 60\degr &   9.33  &   2.38  &   42.6  &    7.8 \\
F6   &  3.9\cdot 10^{-3}  &  16  & 75\degr &   9.78  &   2.23  &   45.5  &    7.6 \\
F7   &  3.9\cdot 10^{-3}  &  16  & 90\degr &   10.2  &   2.40  &   42.2  &    7.4 \\
\hline
G1   &  2.5\cdot 10^{-4}  & 100  &  0\degr &   24.6  &   7.37  &   34.4  &   10.8 \\
G2   &  2.5\cdot 10^{-4}  & 100  & 15\degr &   24.5  &   7.30  &   34.7  &   11.0 \\
G3   &  2.5\cdot 10^{-4}  & 100  & 30\degr &   24.5  &   7.14  &   35.5  &   11.7 \\
G4   &  2.5\cdot 10^{-4}  & 100  & 45\degr &   24.3  &   6.91  &   36.6  &   12.8 \\
G5   &  2.5\cdot 10^{-4}  & 100  & 60\degr &   24.6  &   6.45  &   39.3  &   13.3 \\
G6   &  2.5\cdot 10^{-4}  & 100  & 75\degr &   25.9  &   5.77  &   43.9  &   11.6 \\
G7   &  2.5\cdot 10^{-4}  & 100  & 90\degr &   24.6  &   5.17  &   49.0  &    9.4 \\
\hline
H1   &  3.0\cdot 10^{-5}  & 400  &  0\degr &   51.0  &  17.05  &   29.7  &   15.9 \\
H2   &  3.0\cdot 10^{-5}  & 400  & 15\degr &   51.0  &  16.90  &   30.0  &   16.7 \\
H3   &  3.0\cdot 10^{-5}  & 400  & 30\degr &   50.7  &  16.64  &   30.5  &   18.2 \\
H4   &  3.0\cdot 10^{-5}  & 400  & 45\degr &   50.4  &  16.22  &   31.2  &   19.8 \\
H5   &  3.0\cdot 10^{-5}  & 400  & 60\degr &   50.4  &  15.07  &   33.6  &   20.3 \\
H6   &  3.0\cdot 10^{-5}  & 400  & 75\degr &   49.2  &  13.30  &   38.1  &   17.7 \\
H7   &  3.0\cdot 10^{-5}  & 400  & 90\degr &   45.0  &  12.03  &   42.1  &   14.5 \\
          \hline
          \end{array}
          \vspace{-.2cm}
          $$ \tablefoot{All of the runs have initially 
            $\RaF=5.1\cdot10^{12}$, corresponding to
            $\Fn = 4.6\cdot 10^{-6}$. $\PraSGS = 1$ such that
            $\PeSGS = \Rey$, and $\xi_0 = 0.054$. The grid resolution
            in all runs is $288^3$.}
\end{table}

\begin{figure*}
  \includegraphics[width=\textwidth]{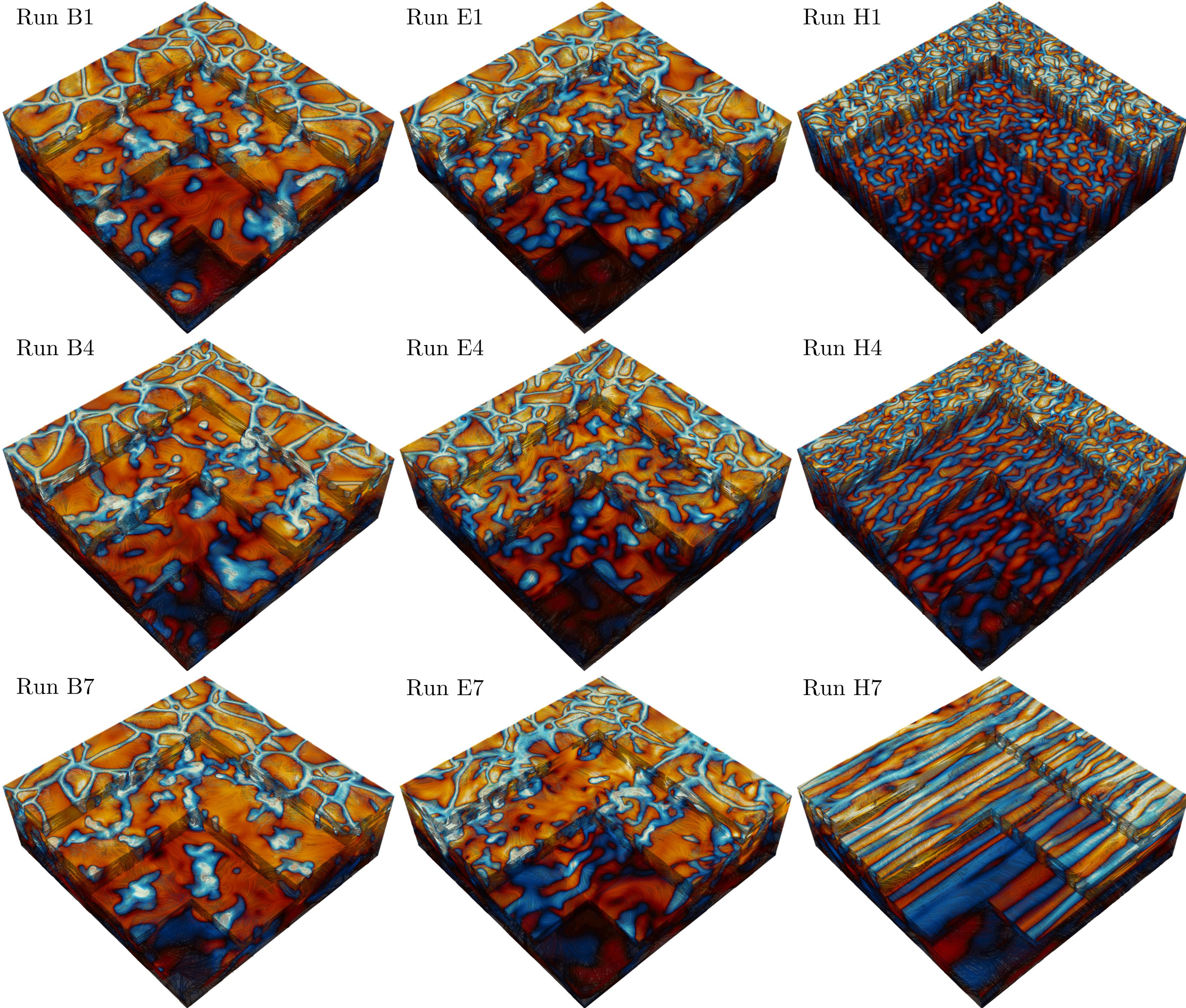}
  \caption{Flow fields from Runs B1, B4, and B7 with
    $\CoF \approx 0.55$ (left column), Runs~E1, E4, and E7 with
    $\CoF \approx 4.5\ldots 4.9$ (middle), and from Runs~H1, H4, and
    H7 with $\CoF \approx 12\ldots 17$ (right) at latitudes
    $\theta = 0\degr$ (top row), $\theta = 45\degr$ (middle), and
    $\theta = 90\degr$ (bottom).}
\label{fig:boxes}
\end{figure*}

\section{Results} \label{sec:results}

Some of the current slowly rotating runs use saturated snapshots of
the progenitor Run~K3h from \cite{Kapyla_2019_AA_631_122} as initial
conditions whereas several sets of runs were run from scratch. In
total, eight sets of runs with different $\CoF$ with seven runs
corresponding to different values of $\theta$ were made; see
\Table{tab:runs1}.

\subsection{Description of the flow fields}

\Figu{fig:boxes} shows representative flow fields from runs with slow
(Set B), intermediate (Set E), and rapid rotation (Set H) highlighting
the transition from cellular convection at slow rotation to columnar
convection at rapid rotation. The effects of rotation are not apparent
in the most slowly rotating cases at any latitude; see the left column
of \Figa{fig:boxes}. For intermediate rotation the average size of
convection cells somewhat reduced and alignment of the convection
cells with the rotation vector is discernible at the equator; see the
middle column of \Figa{fig:boxes}. For rapid rotation the convection
cells are significantly smaller and strongly aligned with the rotation
vector; right column of \Figa{fig:boxes}. This is particularly
apparent at the equator ($\theta= 90\degr$, Run~H7) where structures
are aligned with the $x$ direction. The dominant size of convective
structures at the pole ($\theta=0\degr$) was shown to follow the
Coriolis-Inertial-Archemedean (CIA) scaling $\ell \propto \Co^{-1/2}$
in an earlier study \citep{Kapyla_2024_AA_683_221}. A similar result
was found for the equator ($\theta=90\degr$), corresponding here to
Runs~B7, E7, and H7 shown on the bottom row of \Figa{fig:boxes}, for
spectra taken in the longitudinal ($y$) direction in
\cite{Bekki_2025_AA_703_262}. The evident horizontal anisotropy is a
key ingredient leading to non-zero $\Lambda$ effect.

\begin{figure*}
  \includegraphics[width=\textwidth]{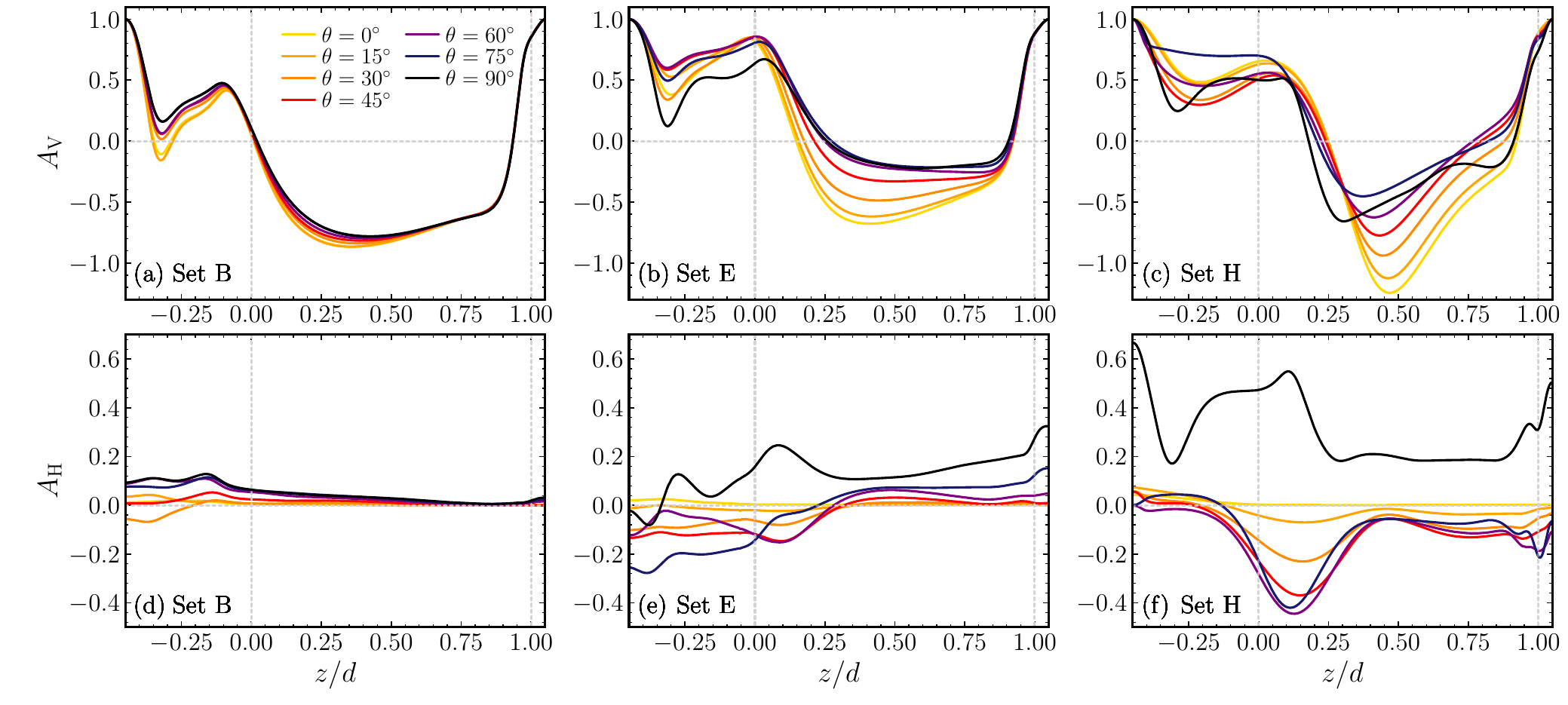}
\caption{Anisotropy parameters $\AAV$ (top row) and $\AAH$ (bottom)
  from runs in Set~B with $\CoF\approx 0.55$ (left), Set~E with
  $\CoF\approx 4.5\ldots 4.9$ (middle), and from Set~H with
  $\CoF\approx 12\ldots17$ (right).}
\label{fig:plot_aniso}
\end{figure*}

\subsection{Anisotropy of turbulence}

The vertical and horizontal $\Lambda$ effects are to lowest order
proportional to the turbulence anisotropies given by $\AAV$ and
$\AAH$, respectively
\citep[e.g.][]{Rudiger_1980_GAFD_16_239}. Representative results for
$\AAV$ and $\AAH$ from runs with slow (Set~B), intermediate (Set~E),
and rapid rotation (Set~H) are shown in \Figa{fig:plot_aniso}. In the
slowly rotating cases $\AAV$ is negative throughout the convection
zone; similarly to non-rotating cases (not shown). For
$\CoF\approx0.55$ (Set~B) the vertical anisotropy $\AAV$ remains
practically constant in the upper part of the convection zone
($z/d \gtrsim 0.65$) as a function of latitude. The absolute value of
$\AAV$ decreases slightly in the lower part of the convection zone
from the pole toward the equator. For intermediate (Set~E) and rapid
rotation (Set~H) the latitudinal variation of $\AAV$ increases; see
panels (b) and (c) of \Figa{fig:plot_aniso}. $\AAV$ generally
decreases from the pole to the equator. In the most rapidly rotating
cases, $\AAV$ at the equator is again somewhat higher than at the low
latitude cases; see panel (c) of \Figa{fig:plot_aniso}. Furthermore,
while $\AAV$ at the pole decreases for intermediate rotation, it is
enhanced in the rapidly rotating cases.

The horizontal anisotropy $\AAH$ is much weaker than $\AAV$ for slow
rotation because it is induced by rotation while the latter is an
inherent property of convection. $\AAH$ vanishes at the pole due to
symmetry and increases monotonically toward the equator for slow and
intermediate rotation; see panels (d) and (e) of
\Figa{fig:plot_aniso}. $\AAH$ is negative for rapid rotation, Set~H;
see panel (f) of \Figa{fig:plot_aniso}, which to lowest order would
imply poleward angular momentum transport that promotes anti-solar
differential rotation
\citep[e.g.][]{Rudiger_1980_GAFD_16_239}. However, the latter result
is valid for slow rotation corresponding to $\Co\ll1$, which is not
satisfied in the current rapidly rotating runs. Furthermore, this
argument relies only on the magnitude of $\AAV$ while spatial scales
of the horizontal flows are also very different in the latitudinal
($x$) and longitudinal ($y$) directions; see \Fig{fig:boxes}.

\begin{figure*}
  \includegraphics[width=\textwidth]{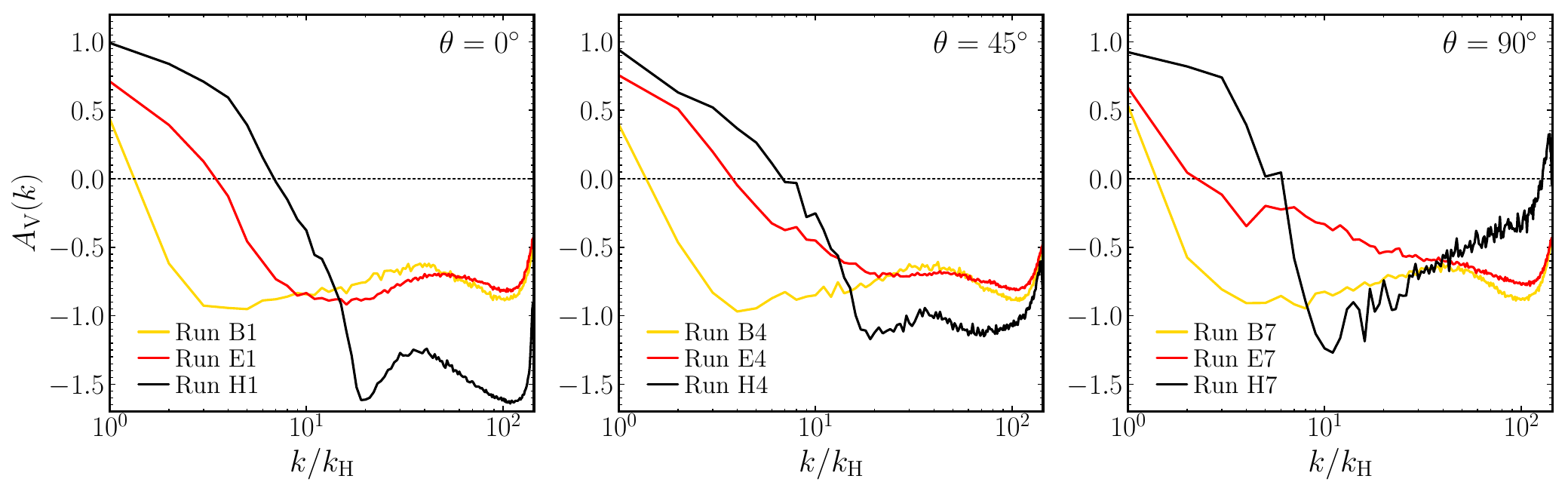}
  \caption{Anisotropy parameter $\AAV(k)$ from runs with slow
    (Runs~B1, B4, and B7), intermediate (Run~E1, E4, and E7), and
    rapid rotation (Run~H1, H4, and H7) at $\theta=0\degr$ (left
    panel), $\theta=45\degr$ (middle), and $\theta=90\degr$ (right),
    respectively, near the middle of the convection zone at
    $z/d = 0.49$.}
\label{fig:paniso_AV}
\end{figure*}

To study the anisotropy in more detail, a spectral decomposition was
made; see \Eqs{equ:AVk}{equ:AHk}. Representative results for $\AAV(k)$
and $\AAH(k)$ are shown in \Figas{fig:paniso_AV}{fig:paniso_AH}. Large
scales are dominated by horizontal flows such that $\AAV(k) > 0$,
similarly as in the non-rotating case (not shown). The zero-crossing
of $\AAV(k)$ occurs at a progressively higher normalized wavenumber
$\tilde{k}=k/k_1$ as the rotation rate increases. The negative
$\AAV(k)$ at intermediate and large wavenumbers are responsible for
$\AAV = \int \AAV(k)dk < 0$ in the bulk of the convection zone in all
cases. For slow rotation (Runs~E[1,4,7]) $\AAV(k)$ does not change
appreciably as a function of $\theta$, whereas for the intermediate
and rapid rotation runs the crossover from positive to negative
$\AAV(k)$ occurs at a lower wavenumber; see the middle and right
panels of \Figa{fig:paniso_AV}.

\begin{figure*}
  \includegraphics[width=\textwidth]{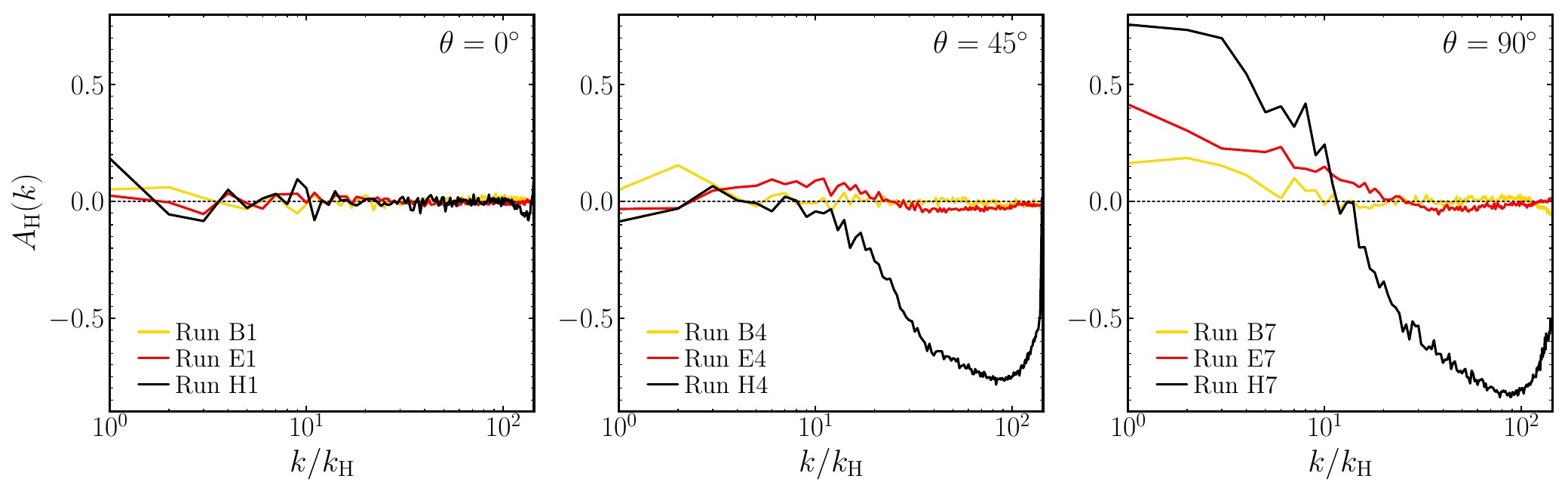}
  \caption{Anisotropy parameter $\AAH(k)$ for the same runs as in
    \Figa{fig:paniso_AV}.}
\label{fig:paniso_AH}
\end{figure*}

As discussed above, the horizontal anisotropy becomes prominent only
at rapid rotation, and it vanishes for $\theta=0\degr$; see
\Figa{fig:paniso_AH}. For slow and intermediate rotation the main
contribution to the overall positive $\AAH$ comes from large and
intermediate scales; see the middle and right panels of
\Figa{fig:paniso_AH}. This can be understood as a consequence of
rotational alignment of the largest convection cells that have the
greatest Coriolis numbers. In the rapid rotation regime a similar
trend is seen at large scales, but the behavior for
$\tilde{k} \gtrsim 10$ the latitudinal ($x$) component dominates
leading to $\AAH = \int \AAH{k}dk < 0$.

Turbulence becomes quasi-two-dimensional for sufficiently rapid
rotation leading to diminishing velocity along $\bm\Omega$
\citep[e.g.][]{Muller_Thiele_2007_EL_77_34003}. Strong density
stratification has a similar effect along the direction of the
gravitational acceleration
\citep[e.g.][]{Boffetta_2023_Atmosphere_14_1688}. The latter has been
adopted as one of the building blocks for theories of angular momentum
transport in stellar convection zones
\citep[e.g.][]{Kitchatinov_Rudiger_1993_AA_276_96}. This approach was
generalized by \cite{Kitchatinov_Rudiger_2005_AN_326_379} who
introduced an anisotropy parameter $a$ that characterizes the
difference between vertical and horizontal motions in the unperturbed
case; see their equation A14. Assuming for simplicity that the scale
ratio $\lcorr^2/L^2 =1$, the vertical anisotropy parameter $\AAV$ can
be written in terms of $a$ as follows:
\begin{eqnarray}
\AAV = \onehalf - \onefifth a,
\end{eqnarray}
such that $a={5 \over 2}$ corresponds to isotropy. The range of $\AAV$
in \Fig{fig:paniso_AV} inside the convection zone is roughly $-0.2$ to
$-1.2$, corresponding to $a = 3.5 \ldots 8.5$ which is significantly
higher than the values typically used in mean-field approaches
\citep[e.g.][]{Kitchatinov_Rudiger_2005_AN_326_379,
  Pipin_Kosovichev_2018_ApJ_854_67}.

\subsection{Reynolds stress and $\Lambda$ effect}

\begin{figure*}
  \includegraphics[width=\textwidth]{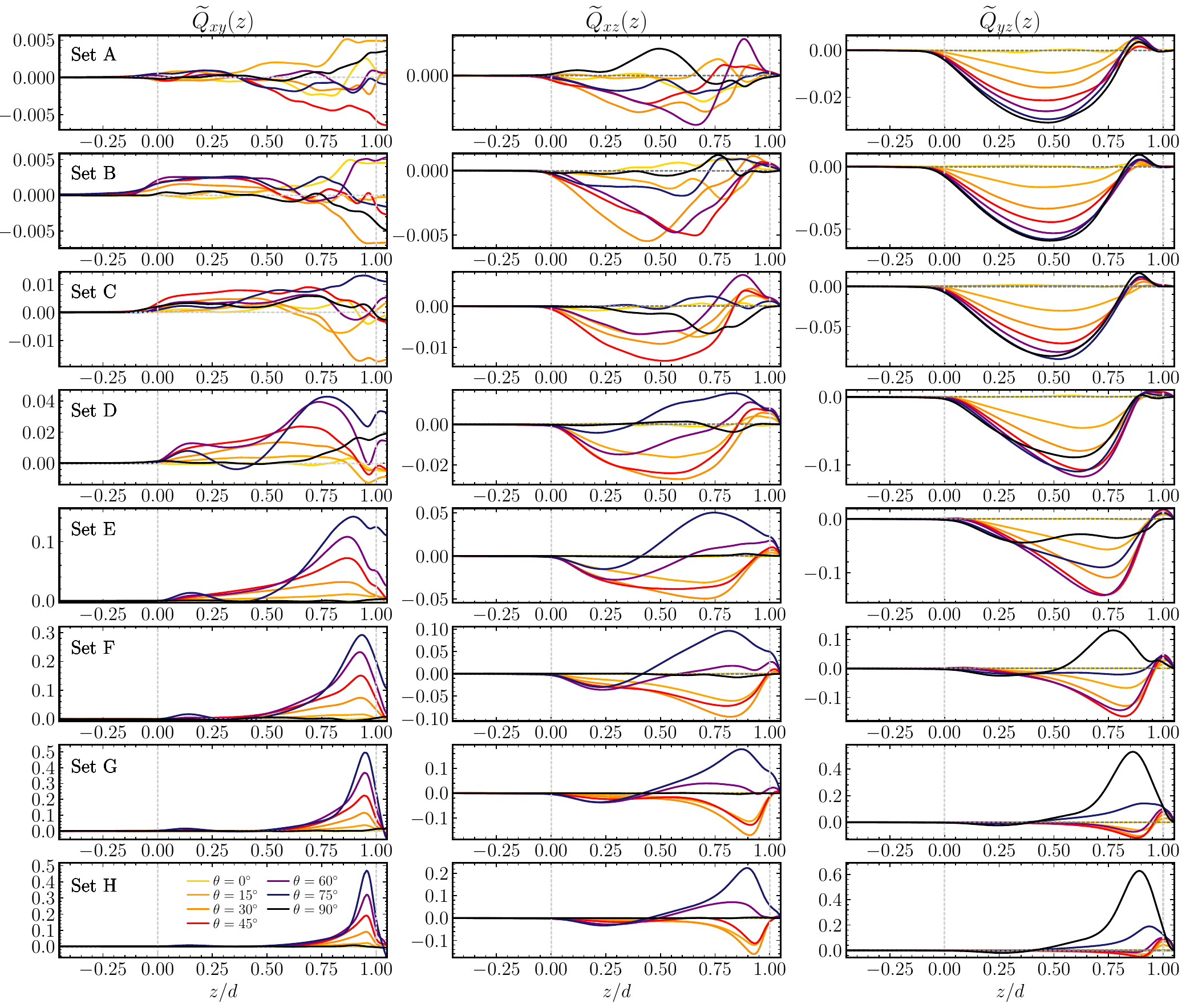}
  \caption{Off-diagonal Reynolds stresses $\tQij{xy}$ (left column),
    $\tQij{xz}$ (middle), and $\tQij{yz}$ (right) from all of the
    runs. The set of runs and the corresponding $\CoF$ are denoted on
    each row on the left panel.}
\label{fig:plot_stress}
\end{figure*}

Horizontally and temporally averaged off-diagonal Reynolds stresses
are shown in \Fig{fig:plot_stress}. The stresses are assumed to be due
to non-diffusive origin, and are attributed here to the $\Lambda$
effect since horizontally averaged mean flows have been suppressed in
the present simulations. The horizontal stress $\qxy$ is responsible
for generating horizontal or latitude-dependent differential rotation
in spherical coordinates. $\qxy$is positive on average almost
everywhere corresponding to equatorward flux of angular momentum; see
the left column of \Fig{fig:plot_stress}. For the slowest rotation
rates the data is not sufficiently converged to rule out negative
values, which however are theoretically expected only in cases where
shear is retained \citep{Rudiger_et_al_2019_AA_630_109}. As the
rotation rate is increased, the horizontal stress is increasingly
concentrated near the equator; see the data for Sets~E to H on the
four bottom left panels of \Fig{fig:plot_stress}. This behavior is
commonly seen in f-plane simulations
\citep[e.g.][]{Chan_2001_ApJ_548_1102, Kapyla_et_al_2004_AA_433_793,
  Hupfer_et_al_2005_AN_326_223} of convection but not in corresponding
forced turbulence calculations
\citep[e.g.][]{Kapyla_Brandenburg_2008_AA_488_9,
  Kapyla_2019_AA_622_195}. The main difference between the forced
turbulence and convection setups is the lack of stratification and
thermodynamics in the former such that, for example, thermal Rossby
waves are not excited. Convective structures corresponding to the
latter are the main contributor to $\qxy$ near the equator \citep[see
also][]{Kapyla_et_al_2011_AA_531_162}. On the other hand, global
convection simulations do not show a similar concentration of
horizontal stress ($\qtp$) near the equator
\citep[e.g.][]{Kapyla_et_al_2011_AA_531_162}. This is likely explained
by the differences in geometry: in the current set-up the horizontal
stress cannot generate a mean flow in the horizontally averaged sense
because of the horizontal periodicity of the domain, whereas in global
simulations the divergence of $\qtp$ leads to a mean flow. However, it
is possible to generate mean flows that depend on $x$ or $y$ but
vanish under a horizontal average even in the present geometry,
although this has not been observed in the current simulations. A
possible reason is that the horizontal size of the domain is too small
or that the horizontal aspect ratio of the domain needs to be unequal
to unity which effectively imposes a preferred direction
\citep[e.g.][]{Guervilly_Hughes_2017_PRF_2_113503,Currie_et_al_2020_MNRAS_493_5233}.
Furthermore, formation of large-scale vortices
\citep[e.g.][]{Chan_2007_AN_328_1059, Kapyla_et_al_2011_ApJ_742_34,
  Guervilly_2014_JFM_758_407} is not observed in the current
simulations most likely because the fluid Reynolds number is too low;
see the discussion in \cite{Kapyla_2024_AA_683_221}.

The meridional stress $\qxz$ does not directly participate in the
angular momentum transport but it has an indirect influence via the
meridional flow. This component is predominantly negative for slow
rotation ($\Co \lesssim 0.25$, Sets~A to C) with a magnitude similar
to $\qxy$. For $\Co\gtrsim 0.6$, $\qxz$ first changes sign for
$\theta=75\degr$ and at even more rapid rotation also for
$\theta=60\degr$. This is in contrast to earlier forced turbulence
simulations which yielded $\qxz<0$ also for rapid rotation
\citep[e.g.][]{Kapyla_2019_AA_622_195}. It is possible that the
difference is also due to the thermal Rossby waves that were absent in
the isothermal forced turbulence models. The magnitude of $\qxz$ is
less than the other off-diagonal components for $\Co\gtrsim 0.6$. The
impact of this stress component for stellar differential rotation has
not been studied in detail.

The vertical stress $\qyz$ generates radial differential rotation.
For slow rotation ($\Co\lesssim 0.6$) $\qyz$ is negative almost
everywhere and roughly proportional to $\Omega$. This agrees with
analytic results and forced turbulence simulations
\citep[e.g.][]{Kapyla_Brandenburg_2008_AA_488_9,Kapyla_2019_AA_622_195}. However,
at sufficiently rapid rotation the sign of $\qyz$ changes near the
equator where large positive values are observed; see the data for
Sets~F to H in the middle column of \Figa{fig:plot_stress}. This
feature is absent in isothermal forced turbulence models and is due to
thermal Rossby waves that are excited in rotating convection
\citep[e.g.][]{Busse_2002_PhFl_14_1301}. Analyses of the radial stress
$\qrp$, corresponding to $\qyz$ in the current Cartesian geometry, in
spherical shells and wedges have shown that the large-scale convective
modes associated with the thermal Rossby waves produce the dominant
contribution to $\qrp$ and produce equatorial acceleration
\citep[e.g.][]{Kapyla_2023_AA_669_98}. This is in contrast to some
hydrodynamic mean-field models where the radial stress vanishes for
rapid rotation and equatorial acceleration is achieved via equatorward
horizontal flux of angular momentum
\citep[e.g.][]{Kitchatinov_2013_IAUS_294_399}.

\begin{figure*}
  \includegraphics[width=\textwidth]{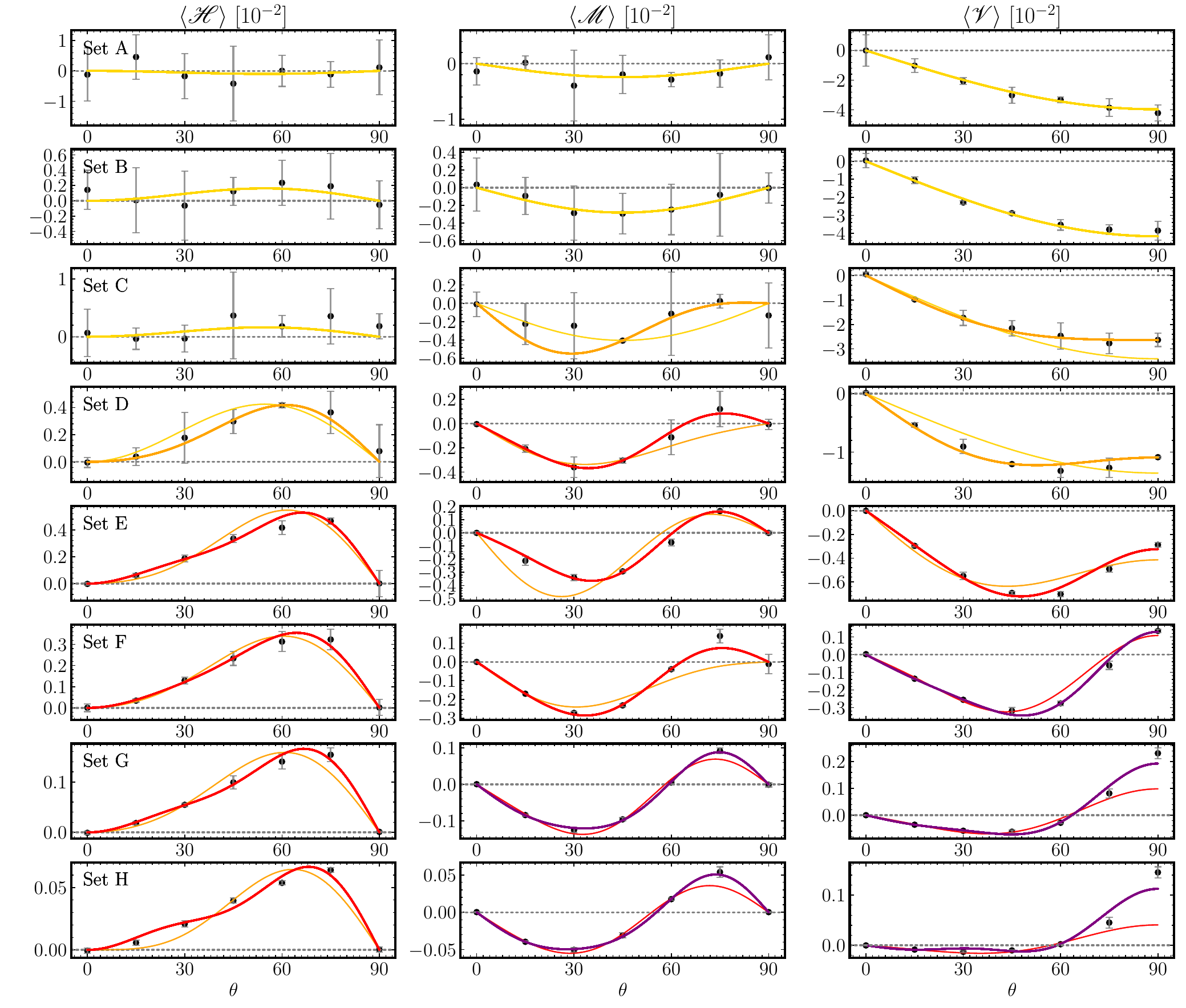}
  \caption{Volume and time-averaged $\Lambda$ coefficients
    $\brac{\mathscr{H}}$ (left panels), $\brac{\mathscr{M}}$ (middle),
    and $\brac{\mathscr{V}}$ (right) from all sets of runs along with
    best fits to \Equsa{equ:Vfit}{equ:Mfit}. The thick lines show an
    adequate fit according to the criteria in the text and the thin
    lines show fits with one fewer coefficient for comparison. The
    colors denote the maximum number of coefficients $\nmax$ taken
    into account in the fits such that $\nmax=0,1,2,3$ correspond to
    yellow, orange, red, and purple, respectively.}
\label{fig:plot_Lambda_vol}
\end{figure*}

The non-diffusive Reynolds stresses from \Equs{equ:Qtp}{equ:Qrp} can
be represented in terms of the $\Lambda$-effect coefficients as
\begin{eqnarray}
  \mathscr{H} &\equiv& H \cos\theta = \twofifteenths \Co^{-1 }\tQij{xy}, \label{equ:HHH} \\
  \mathscr{M} &\equiv& M \sin\theta\cos\theta = \twofifteenths \Co^{-1 }\tQij{xz}, \label{equ:MMM} \\
  \mathscr{V} &\equiv& V \sin\theta = \twofifteenths \Co^{-1 }\tQij{yz}, \label{equ:VVV}
\end{eqnarray}
where $\nutz = {4 \over 15} \urms k_1$ was used, and where the tilde
refers to normalization by $\urms^2$. The functional form of $\nutz$
is taken for a non-rotating case from the analytic study of
\cite[][]{Kitchatinov_et_al_1994_AN_315_157}. Results from numerical
simulations of isotropically forced homogeneous turbulence tend to be
somewhat higher; see \cite{Kapyla_et_al_2020_AA_636_93}. Therefore the
choice of $\nutz$ in the normalization leads to some uncertainty in
the absolute values of the coefficients. We follow the same procedure
as in \cite{Kapyla_2019_AA_622_195}, and expand the coefficients in
powers of $\sin^2\theta$
\begin{eqnarray}
H = \sum_1^{\nmax} H^{(i)} \sin^{2n}\theta, \label{equ:Hfit} \\
M = \sum_0^{\nmax} M^{(i)} \sin^{2n}\theta, \label{equ:Mfit} \\
V = \sum_0^{\nmax} V^{(i)} \sin^{2n}\theta, \label{equ:Vfit}
\end{eqnarray}
where $H^{(0)}$ is assumed to vanish due to symmetry. The Reynolds
stresses vary as a function of height but here we apply volume
averaging, denoted by angle brackets $\brac{.}$, over the range
$0 \le z/d \le 1$ to simplify the analysis. The coefficients $V$, $H$,
and $M$ were fitted with up to $\nmax=3$ using \Equs{equ:HHH}{equ:VVV}
and \Equs{equ:Hfit}{equ:Vfit} with volume averaged numerical data for
the Reynolds stresses. The fit was deemed accurate enough when the
mean error of the mean between the numerical data and the
reconstruction with the fitted coefficients did not decrease by more
then ten per cent when an additional coefficient was taken into
account. However, $\nmax$ higher than 3 would be needed to fit the
vertical stress $\brac{\widetilde{Q}_{yz}}$ for the most rapidly
rotating cases (Sets~G and H) but this was not pursued. The numerical
data as well as the fits are shown in \Figa{fig:plot_Lambda_vol} for
all runs. Furthermore, \Table{tab:Lambdacoefs} summarizes the fitted
coefficients.

\begin{table*}[t!]
\centering
\caption[]{Summary of the fitted $\Lambda$ coefficients.}
  \label{tab:Lambdacoefs}
       \vspace{-0.5cm}
      $$
          \begin{array}{p{0.035\linewidth}ccccccccccc}
          \hline
          \hline
          \noalign{\smallskip}
            Set & H^{(1)} & H^{(2)} & H^{(3)} & M^{(0)} & M^{(1)} & M^{(2)} & M^{(3)} & V^{(0)} & V^{(1)} & V^{(2)} & V^{(3)} \\
            \hline
            A &-0.26 &    -  &   -  & -0.48 &    -  &    -  &   -  & -3.96 &    -  &    -  &   -  \\
            B & 0.42 &    -  &   -  & -0.56 &    -  &    -  &   -  & -4.17 &    -  &    -  &   -  \\
            C & 0.43 &    -  &   -  & -1.72 &  1.81 &    -  &   -  & -3.85 &  1.22 &    -  &   -  \\
            D & 0.39 &  0.96 &   -  & -0.80 & -0.57 &  1.91 &   -  & -2.31 &  1.22 &    -  &   -  \\
            E & 1.16 & -2.02 & 2.99 & -0.62 & -1.47 &  3.02 &   -  & -1.10 & -0.45 &  1.22 &   -  \\
            F & 0.62 & -0.52 & 1.24 & -0.66 & -0.28 &  1.40 &   -  & -0.55 &  0.39 & -1.30 & 1.59 \\
            G & 0.35 & -0.67 & 0.99 & -0.38 &  0.68 & -1.48 & 1.71 & -0.16 &  0.37 & -1.01 & 0.99 \\
            H &  0.18 & -0.44 & 0.56 & -0.18 &  0.42 & -0.79 & 0.84 & -0.05 &  0.34 & -0.89 & 0.72 \\
            \hline
          \end{array}
          \vspace{-.2cm}
          $$
          \tablefoot{Values of all coefficients are boosted by a
            factor of $10^2$. A missing value indicates that the fit
            with fewer coefficients was deemed sufficiently accurate.}
\end{table*}

The horizontal $\Lambda$ effect is positive for all rotation rates,
although in the slowest rotation cases studied (Sets~A, B, and C) the
results are not statistically significant. These results are in
accordance with quasi-linear theory
\citep[e.g.][]{Rudiger_1989_Differential_Rotation_and_Stellar_Convection},
where negative values $\mathscr{H}$ are expected only when large-scale
flows \citep[][]{Rudiger_et_al_2019_AA_630_109} or magnetic fields
\citep[e.g.][]{Kitchatinov_Rudiger_2004_AN_325_496,
  Kapyla_2019_AA_622_195} are present, both of which are absent in the
current simulations. The concentration of the horizontal stress near
the equator is reflected in the horizontal $\Lambda$ effect
coefficient $\mathscr{H}$, although this behavior is subdued in
\Figa{fig:plot_Lambda_vol} due to the volume averaging. Nevertheless,
$\brac{\mathscr{H}}$ tends to have its maximum at the nearest latitude
away from the equator for $\Co > 0.6$ (Sets~E to H). This is reflected
by the need to use coefficients up to $H^{(3)}$ to fit the data
already in Set~E; see the second to fourth columns of
\Table{tab:Lambdacoefs}. More data points at lower latitudes would be
needed to determine the full latitude profile but this is out of the
scope of the present study. In \cite{Kapyla_2019_AA_622_195} all of
the $H^{(i)}$ coefficients were found to be positive which is not the
case for $H^{(1)}$ in the present study.

The volume-averaged meridional $\Lambda$ effect $\brac{\mathscr{M}}$,
shown in the middle column of \Figa{fig:plot_Lambda_vol}, is negative
for slow rotation up to around $\Co = 0.25$ (Set~C). For more rapid
rotation, $\brac{\mathscr{M}}$ changes sign near the equator which was
not observed in the forced turbulence simulations of
\cite{Kapyla_2019_AA_622_195}, but was present in earlier simulations
by
\cite{Kapyla_Brandenburg_2008_AA_488_9}. \cite{Pulkkinen_et_al_1993_AA_267_265}\footnote{Their
  $M^{(i)}$ corresponds to $M^{(i-1)}$ of the current study.} reported
$M^{(0)}=-0.03$ and $M^{(1)}=0.06$, although the quality of their fit
is rather modest; see their Fig.~11. The signs coincide with the
current Set~C although the values are greater than in the current
runs. This is possibly because mean flows were retained in the runs of
\cite{Pulkkinen_et_al_1993_AA_267_265} that also contribute to the
Reynolds stress; see
\citep[e.g.][]{Rudiger_et_al_2019_AA_630_109}. For more rapid rotation
(Sets~D onward), at least three coefficients are needed to fit the
data. The sign of $M^{(0)}$ coincides with that in forced turbulence
simulations of \cite{Kapyla_2019_AA_622_195} but the behavior of the
higher order coefficients is more complex.

The vertical $\Lambda$ effect $\brac{\mathscr{V}}$ is negative for for
$\CoF\lesssim 4.56$ (Sets~A to E). In the two slowest rotating sets A
and B, only the fundamental mode of the $\Lambda$ effect, described by
$V^{(0)}$ is needed to fit the numerical data; see
\Table{tab:Lambdacoefs}. The absolute value of $V^{(0)}$ is of the
order of 0.05 in the slowest rotation cases. The sign and functional
form agree with forced turbulence simulations
\citep[e.g.][]{Kapyla_2019_AA_622_195, Barekat_et_al_2021_AA_655_79}
but the absolute value is an order of magnitude smaller in the current
results although the vertical anisotropy of the flow as measured by
$\LamV$ is similar in both cases; compare \Figa{fig:plot_aniso} with
Fig.~3 of \cite{Barekat_et_al_2021_AA_655_79}. For $\Co\gtrsim 2.5$
(Sets~F to H), the sign of $\brac{\mathscr{V}}$ changes near the
equator. In the current results $V^{(0)}< 0$ and $V^{(3)}> 0$
everywhere, whereas $V^{(1)}>0$ and $V^{(2)}<0$ apart from Set~E; see
the three last columns in \Table{tab:Lambdacoefs}. In
\cite{Kapyla_2019_AA_622_195}, $V^{(0)}$ is also predominantly
negative, whereas $V^{(1)}<0$ and $V^{(2)}>0$ in contrast to the
current results. Analytic studies do not typically consider
coefficients higher than $V^{(1)}$; see, for example,
\cite{Kitchatinov_Rudiger_2005_AN_326_379,
  Pipin_Kosovichev_2018_ApJ_854_67}.

The qualitative differences between the analytic theories and
isothermal forced turbulence simulations in comparison to the current
results for the vertical $\Lambda$ effect are likely due to the
dominating influence of thermal Rossby waves that appear as tilted
columnar convection cells near the equator
\citep[e.g.][]{Kapyla_2023_AA_669_98}. Thermal Rossby waves are
typically not captured by analytics that rely on simplified turbulence
models and therefore a positive $\brac{\mathscr{V}}$ for rapid
rotation is absent in such theories. This applies also to analytic
theories where the convective heat flux is taken into account
\citep[e.g.][]{Kleeorin_Rogachevskii_2006_PRE_73_046303}. However,
\cite{Pipin_Kosovichev_2018_ApJ_854_67} found that the standard
mean-field theory does yields a sign change of the vertical $\Lambda$
effect for rapid rotation although the physical interpretation of this
effect is currently unclear.

\section{Conclusions}
\label{sec:conclusions}

The current simulations show that the radial non-diffusive angular
momentum transport, parameterized by the vertical $\Lambda$ effect, is
negative for slow rotation ($\Co \lesssim 1.3$) in rotating
density-stratified convection. The horizontal $\Lambda$ effect is
positive, corresponding to equatorward angular momentum flux, for all
rotation rates. These findings are in qualitative agreement with
isothermal forced turbulence simulations
\citep[e.g.][]{Kapyla_Brandenburg_2008_AA_488_9,
  Kapyla_2019_AA_622_195, Barekat_et_al_2021_AA_655_79} and analytic
mean-field theories
\citep[e.g.][]{Kitchatinov_Rudiger_1995_AA_299_446,
  Kleeorin_Rogachevskii_2006_PRE_73_046303}. For more rapid rotation
the radial $\Lambda$ effect changes sign and the horizontal flux is
increasingly concentrated at low latitudes. The change of sign of of
the vertical $\Lambda$ effect is due to the emergence of prograde
propagating thermal Rossby waves which are not accounted for in
analytic theories of angular momentum transport
\citep[e.g.][]{Rudiger_1989_Differential_Rotation_and_Stellar_Convection,
  Kitchatinov_Rudiger_2005_AN_326_379}. In global convection
simulations the thermal Rossby waves are the cause of solar-like
differential rotation. On the other hand, thermal Rossby waves have
not been observed in the Sun \citep[e.g.][and references
therein]{Birch_et_al_2024_PhFl_36_117136}, and if they are present,
they must be much weaker than in simulations. Therefore the origin of
solar and stellar differential rotation is related to the ongoing
debate regarding the nature of convection in the Sun \citep[e.g.][and
references therein]{Schumacher_Sreenivasan_2020_RvMP_92_041001,
  Hotta_et_al_2023_SSRv_219_77}, often referred to as convective
conundrum \citep[e.g.][]{OMara_et_al_2016_AdSpR_58_1475}.

The current study probes the hydrodynamic regime with simulations
where the density stratification is much weaker than in stars and
where the transition to optically thin photosphere is modeled rather
crudely. Magnetic fields have been shown to strongly affect the
angular momentum transport and large-scale flows in global simulations
\citep[e.g.][]{Hotta_et_al_2022_ApJ_933_199, Hotta_2025_ApJ_985_163,
  Kapyla_2023_AA_669_98,
  Soderlund_et_al_2025_MNRAS_541_1816}. Magnetic fields have also been
shown to significantly influence the $\Lambda$ effect in theoretical
and idealized numerical studies
\citep[e.g.][]{Kitchatinov_Rudiger_2004_AN_325_496,
  Kapyla_2019_AA_622_195, Kapyla_2019_AN_340_744}. The inability to
incorporate a realistic stratification or proper radiative surface
possibly inhibit non-local entropy rain
\citep[e.g.][]{Spruit_1997_MEMSAI_68_397, Brandenburg_2016_ApJ_832_6,
  Kapyla_2025_AA_698_L13} in the current studies. The effects to
convective angular momentum transport of these aspects need to be
addressed in future studies.

\begin{acknowledgements}
  The simulations were performed using the resources granted by the
  Gauss Center for Supercomputing for the Large-Scale computing
  project ``Cracking the Convective Conundrum'' in the Leibniz
  Supercomputing Centre's SuperMUC-NG supercomputer in Garching,
  Germany. This work was supported in part by the Deutsche
  Forschungsgemeinschaft Heisenberg programme (grant No.\ KA
  4825/4-1).
\end{acknowledgements}

\bibliographystyle{aa}
\bibliography{paper}

\end{document}